\documentclass[aps,prb,floatfix,twocolumn,superscriptaddress]{revtex4-1}
\pdfoutput=1
\usepackage{amsmath}
\usepackage{amssymb}
\usepackage{amstext}
\usepackage{amsopn}
\usepackage{amsfonts}
\usepackage{amsxtra}
\usepackage{color}
\usepackage{dcolumn}
\usepackage{graphicx}
\usepackage{hyperref}
\usepackage{bm}
\usepackage{pifont}

\begin{document}

\bibliographystyle{apsrev}

\preprint{Draft version, not for distribution}

%
%

\title[scratekk]{The fate of quasiparticles in the superconducting state}

%
%
%
\author{S.V. Dordevic}
\email{dsasa@uakron.edu}%
\affiliation{Department of Physics, The University of Akron,
Akron, Ohio 44325 USA}%
\affiliation{Departement de Physique de la Matiere Condensee,
Universite de Geneve, Quai Ernest-Ansermet 24, 1211 Geneve 4, Switzerland}%
\author{D. van der Marel}
\affiliation{Departement de Physique de la Matiere Condensee,
Universite de Geneve, Quai Ernest-Ansermet 24, 1211 Geneve 4, Switzerland}%
\author{C.C. Homes}
\affiliation{Condensed Matter Physics and Materials Science Department,
Brookhaven National Laboratory, Upton, New York 11973, USA}%

\date{\today}

%
%
\begin{abstract}
Quasiparticle properties in the superconducting state are masked by
the superfluid and are not directly accessible to infrared
spectroscopy. We show how one can use a Kramers--Kronig
transformation to separate the quasiparticle from superfluid
response and extract intrinsic quasiparticle properties in the
superconducting state. We also address the issue of a narrow
quasiparticle peak observed in microwave measurements, and
demonstrate how it can be combined with infrared measurements to
obtain unified picture of electrodynamic properties of cuprate
superconductors.
\end{abstract}

%
%
%
%
%
\pacs{78.30.Er, 74.72.-h, 74.70.Xa}

\maketitle

\section{Introduction}

\begin{figure}[t]
\vspace*{-0.5cm}%
\centerline{\includegraphics[width=9.5cm]{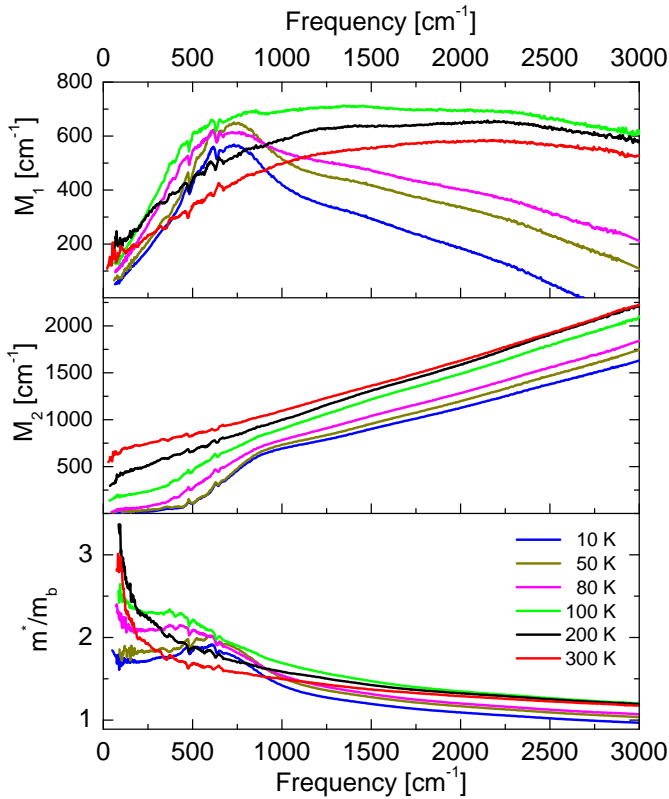}}%
\vspace*{-0.5cm}%
\caption{(Color online). Extended Drude analysis of optimally doped
Bi2212 with T$_c$=~92~K. (a) Real part of memory function M$_1(\omega)$ obtained
from Eq.~\ref{eq:memory}. (b) Imaginary part of memory function M$_2(\omega)$
from Eq.~\ref{eq:memory}. (c) Quasiparticle effective mass $m^{*}(\omega)/m_b$
calculated from Eq.~\ref{eq:mass}. The values of plasma frequency $\omega_p$
used for each temperature are shown in Fig.~\ref{fig:sigma}(c).}
\vspace*{-0.5cm}%
\label{fig:bisco}
\end{figure}

Infrared spectroscopy has in the past several decades become one of
the premier experimental tools in condensed matter physics
\cite{dressel-book,basov11}. Thanks to its versatility, it
has been successfully applied to essentially all types of
condensed matter systems, such as superconductors, topological
insulators, graphene, etc. In particular, in high-T$_c$ cuprate
superconductors, infrared based techniques have been extensively
used to probe a variety of unusual and yet unresolved issues
concerning their unconventional pairing state \cite{basov05}. In
recent years a number of attempts has been made to elucidate the
properties of quasiparticles and their relaxation in the cuprates
\cite{iman13,homes13}. These attempts are based on the so-called
extended-Drude model, which allows both the quasiparticle
scattering rate and their effective mass to acquire frequency
dependence. These two quantities can be straightforwardly obtained
from the complex optical conductivity $\tilde\sigma(\omega)=
\sigma_1(\omega) + i \sigma_2(\omega)$ as:

\begin{equation}
\frac{1}{\tau(\omega)} =\frac{\omega_{p}^{2}}{4\pi}
\Re\left[\frac{1}{\tilde\sigma(\omega)}\right]=
\frac{\omega_{p}^{2}} {4 \pi} \frac{\sigma_1(\omega)}
{\sigma_1^2(\omega)+\sigma_2^2(\omega)} %
\label{eq:tau}
\end{equation}

\begin{equation}
\frac{m^{*}(\omega)}{m_b} =\frac{\omega_{p}^{2}}{4 \pi}
\Im\left[\frac{1}{\tilde\sigma(\omega)}\right] \frac{1}{\omega} =
\frac{\omega_{p}^{2}}{4 \pi} \frac{\sigma_2(\omega)}
{\sigma_1^2(\omega)+\sigma_2^2(\omega)}\frac{1}{\omega}
\label{eq:mass}
\end{equation}
where the plasma frequency $\omega_p^2=4 \pi e^2 n/m_b$ is usually
obtained from the integration of $\sigma_{1}(\omega)$ up to the
frequency of the onset of interband absorption:

\begin{equation}
\omega_p^2= 8 \int_{0+}^{\Omega} \sigma_1(\omega) d \omega %
\label{eq:plasma}
\end{equation}
Equations \ref{eq:tau} and \ref{eq:mass} are the basis of a
so-called {\it one-component} approach \cite{tanner92} for the
interpretation of optical properties, which assumes the existence
of a single type of charge carriers in the system. Closely related
quantities are the optical self-energy
$\tilde{\Sigma}^{opt}$($\omega$) \cite{timusk04} and memory
function $\tilde{M}$($\omega$) \cite{gotze72,iman13} defined as:

\begin{equation}
\tilde{M}(\omega)= 2 \tilde{\Sigma}^{opt}(\omega) =
\omega (\frac{m^*(\omega)}{m_b} - 1) + i \frac{1}{\tau(\omega)}  %
\label{eq:memory}
\end{equation}
As an example, in Fig.~\ref{fig:bisco} we displays the real and
imaginary parts of the memory function $\tilde{M}(\omega)$, as well
as the effective mass $m^{*}(\omega)/m_b$ for optimally doped
Bi$_2$Sr$_2$CaCu$_2$O$_{8+\delta}$ (Bi2212) with T$_c$=~92~K
\cite{tu02}, both in the normal and superconducting state.
Application of Eqs.~\ref{eq:tau}, \ref{eq:mass} and \ref{eq:memory}
to the data in the superconducting state is highly problematic,
even though it has been routinely done. Namely, in the
superconducting state normal fluid coexists with the superfluid,
and the one-component assumption is clearly violated. As was most
recently pointed out by Homes et al. \cite{homes13} this procedure
cannot be used to make any reliable statements about quasiparticle
properties in the superconducting state. In the superconducting
state the response at microwave and far-infrared frequencies is
dominated by the superfluid, causing $\sigma_2(\omega) \gg
\sigma_1(\omega)$ (Ref.~\onlinecite{superfluid-comm}) and it
follows from Eq.~\ref{eq:tau} that 1/$\tau(\omega)$ = M$_2(\omega)$
$\sim \sigma_1 / \sigma_2^2$ acquires small values (see
Fig.~\ref{fig:bisco}). Similarly, indiscriminate application of
Eq.~\ref{eq:mass} to the optical conductivity in the superconducting
state will result in $m^*(\omega)/m_b \sim 1 / (\sigma_2 \omega)$
which decreases when the superfluid forms, and $\sigma_2(\omega)$
increases (see Fig.~\ref{fig:bisco}).


In this work we show how to circumvent this problem, and access
intrinsic quasiparticle properties in the superconducting state.
The procedure expands the range of applicability of one-component
approach, and further extends the power of infrared spectroscopy.
Our procedure is based on Kramers-Kronig transformation of the optical
conductivity. In the next section (Sec.~\ref{sec:kk}) we first
present the formalism in details. In Sec.~\ref{sec:model} we apply
it to model data, which serves to illustrate the main idea and
demonstrate its usefulness. The procedure is then applied to
existing infrared data on optimally doped high-T$_c$ superconductor
Bi2212 with T$_c$= 92 K (Sec.~\ref{sec:bi2212}). We present and
discuss the quasiparticle properties obtained for the first time
below T$_c$. In this section we also address the issue of a narrow
quasiparticle peak that has been observed in microwave
spectroscopy, and show how it can be combined with infrared data to
obtain a unified and self-consistent picture of electrodynamic
properties of Bi2212. Finally, in Section \ref{sec:summary} we
summarize the most important findings made possible by the new
procedure.

\section{Kramer--Kronig approach}
\label{sec:kk}

In this section we present the details of our procedure. The
approach is based on the Kramers-Kronig transformation of optical
conductivity, which we introduced previously to extract the
superfluid density from infrared data \cite{dordevic02}. In the
superconducting state the real part of optical conductivity can be
written as:

\begin{equation}
\sigma_1^{SC}(\omega)= \rho_s \delta(\omega)+\sigma_1^{qp}(\omega)
\label{eq:s1}
\end{equation}
where the first term on the right-hand side is the superconducting,
and the second the quasiparticle contribution. The corresponding
imaginary part of $\sigma_1^{SC}(\omega)$ follows from a
Kramers-Kronig relation:

\begin{equation}
\sigma_2^{SC}(\omega)= \rho_s \frac{1}{\omega}+\sigma_2^{qp}(\omega)
\label{eq:s2}
\end{equation}
where $\rho_s = \omega_s^2$ is the superfluid density or stiffness, and
$\omega_s$ is the superconducting plasma frequency. The Dirac delta
function in $\sigma_1^{SC}(\omega)$ is not accessible in optical
data, which typically start at several meV. However, the 1/$\omega$
term in $\sigma_2^{SC}(\omega)$ is mixed up with
$\sigma_2^{qp}(\omega)$ and contributes to both 1/$\tau$($\omega$)
and m$^*$($\omega$)/m$_b$ (Eqs.~\ref{eq:tau} and \ref{eq:mass}). To
determine the intrinsic quasiparticle properties we must separate
the two terms in Eq.~\ref{eq:s2}. To that end we employ a
Kramers-Kronig transformation on $\sigma_1^{qp}(\omega)$:

\begin{equation}
\sigma_2^{qp}(\omega)=-\frac{2 \omega}{\pi}\int_{0^+}^{\infty}
\frac{\sigma_1^{qp}(\omega')}{\omega'^2-\omega^2}d\omega'.
\label{eq:s2kk}
\end{equation}
We emphasize that this step is completely model-independent; no a
priori assumptions are made about the form of quasiparticle
conductivity. Once $\sigma_2^{qp}(\omega)$ is calculated from
Eq.~\ref{eq:s2kk}, one can calculate the intrinsic scattering rate
and effective mass in the superconducting state (Eqs.~\ref{eq:tau}
and \ref{eq:mass}) using the Kramers-Kronig-corrected
$\sigma_2^{qp}(\omega)$, instead of $\sigma_2^{SC}(\omega)$. Note
that $\sigma_1^{SC}(\omega)$ does not need to be corrected, as the
delta function (Eq.~\ref{eq:s1}) is not accessible to optical
experiments. Using the procedure described above, we can also
calculate the superfluid density from Eq.~\ref{eq:s2}, as was done
before \cite{dordevic02}:

\begin{equation}
\rho_s = \omega_s^2 = \omega ( \sigma_2^{SC}(\omega)-\sigma_2^{qp}(\omega) ).
\label{eq:rhos}
\end{equation}

\section{Model calculations}
\label{sec:model}

\begin{figure}[t]
\vspace*{0.0cm}%
\centerline{\includegraphics[width=9.5cm]{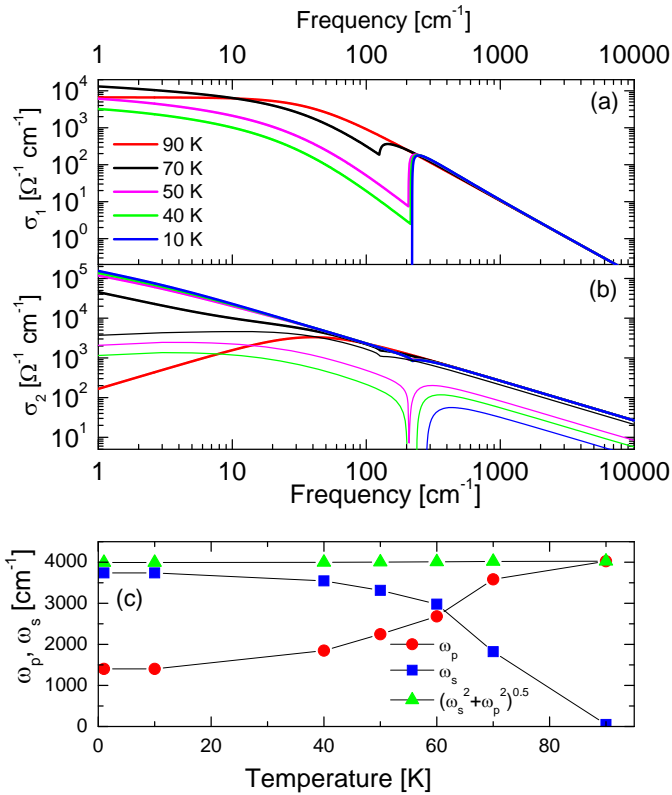}}%
\vspace*{0.0cm}%
\caption{(Color online). (a) Real part of the optical conductivity
$\sigma_1(\omega)$ of a BCS model with T$_c$= 90 K. (b) Imaginary part of the
optical conductivity $\sigma_2(\omega)$. Thick lines represent the model function,
whereas thin lines of the same color represent Kramers-Kronig-corrected $\sigma_2^{qp}(\omega)$.
(c) Temperature dependence of plasma frequency $\omega_p$, superconducting plasma frequency
$\omega_s$ and the total plasma frequency ($\omega_p^2$+$\omega_s^2$)$^{1/2}$.}
\vspace*{0.0cm}%
\label{fig:model}
\end{figure}

To test the procedure and to demonstrate its usefulness in this
section we perform the calculations on model data. We adopt a Drude
model for the normal state, and a BCS model for an s-wave
superconductor in the superconducting state. The model also
includes a quasiparticle peak inside the superconducting gap. The
critical temperature was set at T$_c$= 90 K, and the corresponding
T=0 energy gap is 2$\Delta$= 220~cm$^{-1}$ (27.3~meV). Real and
imaginary parts of $\tilde\sigma(\omega)$ are shown with thick
lines at several temperatures in Fig.~\ref{fig:model}(a) and (b)
respectively. In the superconducting state a characteristic
suppression of $\sigma_1(\omega)$ is observed below the gap. The
spectral weight removed from these frequencies is transferred to
the delta function at zero frequency (Eq.~\ref{eq:s1}). The values
of plasma frequency are calculated from Eq.~\ref{eq:plasma} and
displayed in Fig.~\ref{fig:model}(c) with red circles. Note that in
Eq.~\ref{eq:plasma} the integral starts from 0+, which emphasizes
the fact that in the superconducting state only the quasiparticle
contribution should be counted towards $\omega_p$. On the other
hand, $\sigma_2(\omega)$ is dominated by a characteristic
1/$\omega$ response of the superfluid (Eq.~\ref{eq:s2}). We also
note that the absolute values of $\sigma_2(\omega)$ are at least an
order of magnitude larger than $\sigma_1(\omega)$.


We now apply the procedure outlined in the previous section. In
Fig.~\ref{fig:model}(b) with thin lines we display
$\sigma_2^{qp}(\omega)$ calculated from Eq.~\ref{eq:s2kk}. We note
that these Kramers-Kronig-corrected curves are not dominated by
1/$\omega$ superfluid response, but instead display a broad peak at
finite frequencies, similar to the one seen at 90 K. The removal of
superfluid response also reveals pronounced structure at the gap
frequency, which is not observable in $\sigma_2^{SC}(\omega)$
(before the correction).

\begin{figure}[t]
\vspace*{0.0cm}%
\centerline{\includegraphics[width=9.5cm]{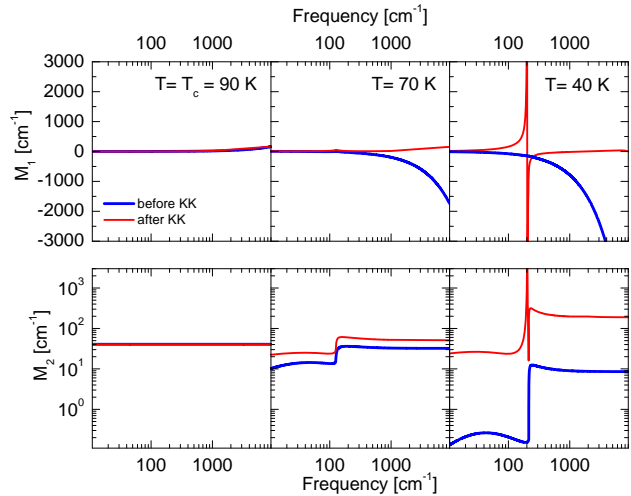}}%
\vspace*{-0.5cm}%
\caption{(Color online). Real and imaginary parts of the memory
function $M_1(\omega)$ and $M_2(\omega)$ for the model shown
in Fig.~\ref{fig:model}. The results are shown before (blue lines)
and after (red lines) Kramers-Kronig correction.}
\vspace*{0.0cm}%
\label{fig:modelM}
\end{figure}

In Fig.~\ref{fig:modelM} we display with blue lines the results for
the real (top panels) and imaginary (bottom panels) parts of the
memory function $\tilde{M}(\omega)$ from Eq.~\ref{eq:memory}. In
the normal state both parts display constant values, typical of the
Drude model. On the other hand, in the superconducting state,
M$_1$($\omega$) is suppressed at higher frequencies, but does not
show any characteristic features at the gap, because the response
is dominated by the superfluid. The imaginary part M$_2$($\omega$)
displays characteristic suppression, especially below the gap. In
the same figure we also plot with red lines the results for the
memory function obtained with Kramers-Kronig-corrected
$\sigma_2^{qp}(\omega)$. Expectedly, in the normal state the memory
function is the same as before. However, in the superconducting
state, the removal of the superfluid contribution reveals a very
pronounced structure at the gap frequency in M$_1(\omega)$. The
imaginary part M$_2(\omega)$ also displays structure at the gap
frequency, but more importantly the suppression below the gap is
much smaller then before. Above the gap M$_2(\omega)$ is enhanced
compared to the normal state.

In Fig.~\ref{fig:model}(c) we display the temperature dependence of
the plasma frequency $\omega_p$ (red circles) obtained from
Eq.~\ref{eq:plasma}, and the superfluid density (blue squares)
obtained from Eq.~\ref{eq:rhos}. As discussed above, the spectral
weight removed form finite frequencies is transferred to the delta
function at zero energy, but the total spectral weight must be
conserved. This is indeed confirmed by Fig.~\ref{fig:model}(c)
where the total (i.e. combined) plasma frequency
($\omega_p^2$+$\omega_s^2$)$^{1/2}$, is shown with green triangles
and is constant within the error bars of numerical calculations.
Note that the application of Eq.~\ref{eq:rhos} to the normal state
data results in a small, but finite value of superconducting plasma
frequency. This is due to numerical errors, and does not imply the
existence of superfluid in the normal state, above T$_c$.


\section{Bi2212}
\label{sec:bi2212}

Before applying the procedure to Bi2212, we must address the issue
of a very narrow quasiparticle peak that has been observed in
microwave measurements \cite{bonn93,shibauchi96,lee96}. Its width
is typically a few meV \cite{bonn93,shibauchi96,lee96}, which is
outside the frequency window of typical infrared measurements. The
existence of this quasiparticle peak is usually ignored during
analysis of infrared data, as it does not contain a lot of spectral
weight and does not significantly affect the calculation of normal
state plasma frequency (Eq.~\ref{eq:plasma}). However, this narrow
peak can produce a significant 1/$\omega$ contribution to
$\sigma_2(\omega)$, which mimics the superfluid response.
If we want to calculate the London penetration depth or the superfluid
stiffness, it must be separated from the superconducting contribution.
We show below that in the case of optimally doped Bi2212 this can
lead to correction of superconducting plasma frequency $\omega_s$
by as much as 40~$\%$.

\begin{figure}[t]
\vspace*{0.0cm}%
\centerline{\includegraphics[width=9.5cm]{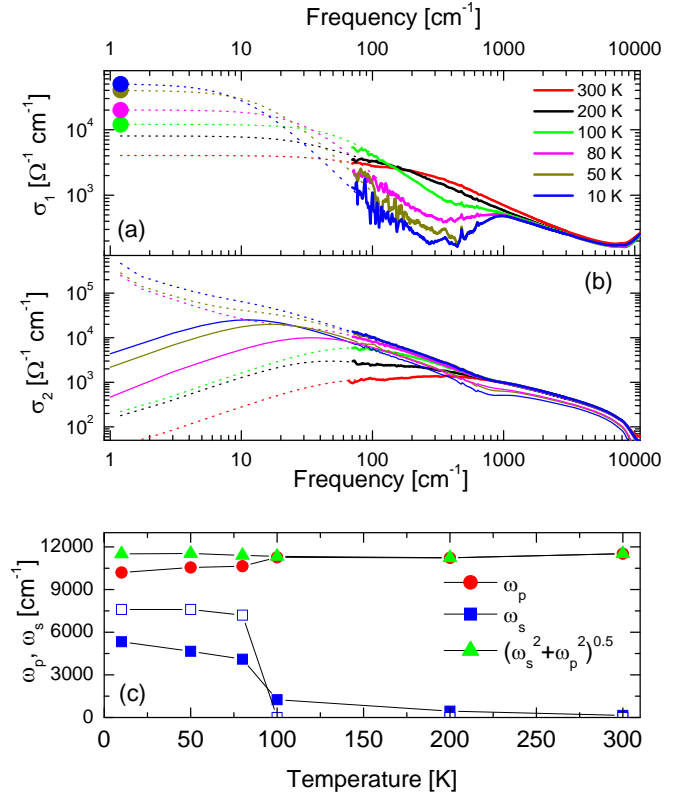}}%
\vspace*{0.0cm}%
\caption{(Color online). (a) Real part of the optical conductivity
$\sigma_1(\omega)$ of optimally
doped Bi2212 with T$_c$= 92 K. Thick lines represent the experimental data,
thin dashed lines of the same color represent Drude extrapolations, and the
circles represent microwave data at 34.7~GHz \cite{lee96}. (b) Imaginary part of the
optical conductivity $\sigma_2(\omega)$. Thick lines represent the experimental
data, thin dashed lines of the same color represent Drude extrapolations and
thin lines represent Kramers-Kronig-corrected $\sigma_2^{qp}(\omega)$. (c) Temperature
dependence of plasma frequency $\omega_p$, superconducting plasma frequency $\omega_s$ and
the total plasma frequency ($\omega_p^2$+$\omega_s^2$)$^{1/2}$.
Open squares represent the values of superconducting plasma
frequency $\omega_s$ obtained without Kramers-Kronig-correction.}
\vspace*{0.0cm}%
\label{fig:sigma}
\end{figure}

In order to perform the integration in Eq.~\ref{eq:s2kk} optical
conductivity data must be extended down to zero frequncy. It was
recently shown that because of the Kramers-Kronig relations between
$\sigma_1(\omega)$ and $\sigma_2(\omega)$ one can calculate the
spectral weight that is located below the lowest measured frequency
\cite{kuzmenko07}. However, the optical functions themselves
($\sigma_1(\omega)$ and $\sigma_2(\omega)$) cannot be retrieved
without making some model assumptions about the optical spectrum.
Here we will make the reasonable assumption that the quasiparticle
contribution can be approximated with the Drude model, and we
combine it with the microwave data on Bi2212. Fig.~\ref{fig:sigma}
displays $\sigma_1(\omega)$ from infrared, as well as the microwave
values at 34.7~GHz $\simeq$ 1.15~cm$^{-1}$ (the values at 14.4 and
24.6~GHz are similar) \cite{lee96}. We now fit the complex
conductivity $\tilde\sigma(\omega)$ {\it simultaneously} with
microwave data, imposing the constraint that the total spectral
weight in the superconducting state is conserved \cite{comm-dirk}.
In Fig.~\ref{fig:sigma} the infrared data is shown with thick
lines, and model fits with dotted lines. The microwave values for
the corresponding temperatures are shown with the circles of the
same color.


With optical conductivity extended down to zero frequency we can
now apply the procedure to Bi2212 and the results are shown in
Fig.~\ref{fig:sigma}(b) with thin lines (only below T$_c$). We can
see that instead of a characteristic 1/$\omega$ divergence, the
spectra display a finite frequency peak, characteristic of
quasiparticle response. The values of normal state plasma frequency
$\omega_p$ and superconducting plasma frequency $\omega_s$ are
shown in Fig.~\ref{fig:sigma}(c) with red circles and blue squares,
respectively. The total plasma frequency
($\omega_p^2$+$\omega_s^2$)$^{1/2}$ (green triangles) is within
1.5~$\%$ of the normal state value. Also shown with empty squares
is the superconducting plasma frequency calculated using
un-corrected $\sigma_2(\omega)$, which can be as much as 40~$\%$
higher than the corrected one.

\begin{figure}[t]
\vspace*{0.0cm}%
\centerline{\includegraphics[width=9.5cm]{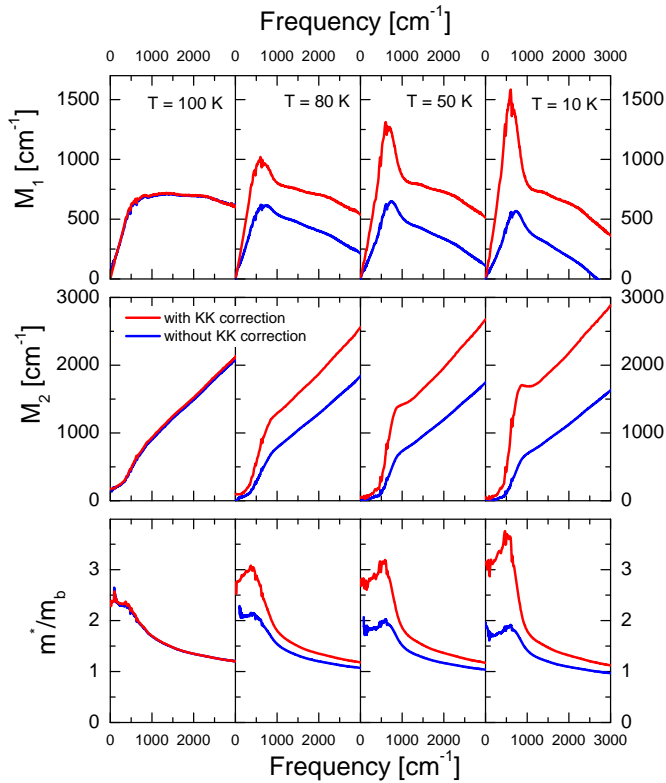}}%
\vspace*{0.0cm}%
\caption{(Color online). Real and imaginary parts of the memory
function $M_1(\omega)$ (top panels) and $M_2(\omega)$ (middle panels),
as well as quasiparticle effective mass m$^*$/m$_b$ (bottom panels)
for the Bi2212 data shown
in Fig.~\ref{fig:sigma}. The results are shown before (blue lines)
and after (red lines) Kramers-Kronig correction.}
\vspace*{-0.5cm}%
\label{fig:kkbisco}
\end{figure}

Once the superfluid contribution is removed from
$\tilde\sigma(\omega)$, one can calculate intrinsic quasiparticle
properties (Eqs.~\ref{eq:tau}, \ref{eq:mass} and \ref{eq:memory}).
Fig.~\ref{fig:kkbisco} displays M$_1$($\omega$) (top panels),
M$_2$($\omega$)=1/$\tau(\omega)$ (middle panels) and effective mass
$m^{*}(\omega)/m_b$ (bottom panels), both before (blue lines) and
after Kramers-Kronig corrections (red lines). Several selected
temperatures are shown, both in the normal (100 K) and
superconducting state (80, 50 and 10 K). Expectedly, optical
functions in the normal state are the same before and after
Kramers-Kronig correction. In the superconducting state, on the
other hand, the corrections are significant and cannot be
neglected. We note that even though the plasma frequency $\omega_p$
decerases in the superconducting state (Fig.~\ref{fig:sigma}(c)),
all optical functions are {\it enhanced} compared to their
un-corrected values. In addition, the structure at around
700~cm$^{-1}$ is much more pronounced in all corrected spectra. The
reason is the removal of superconducting contribution from
$\sigma_2(\omega)$, which exposes the true quasiparticle
properties.


\begin{figure}[t]
\vspace*{0.0cm}%
\centerline{\includegraphics[width=9.5cm]{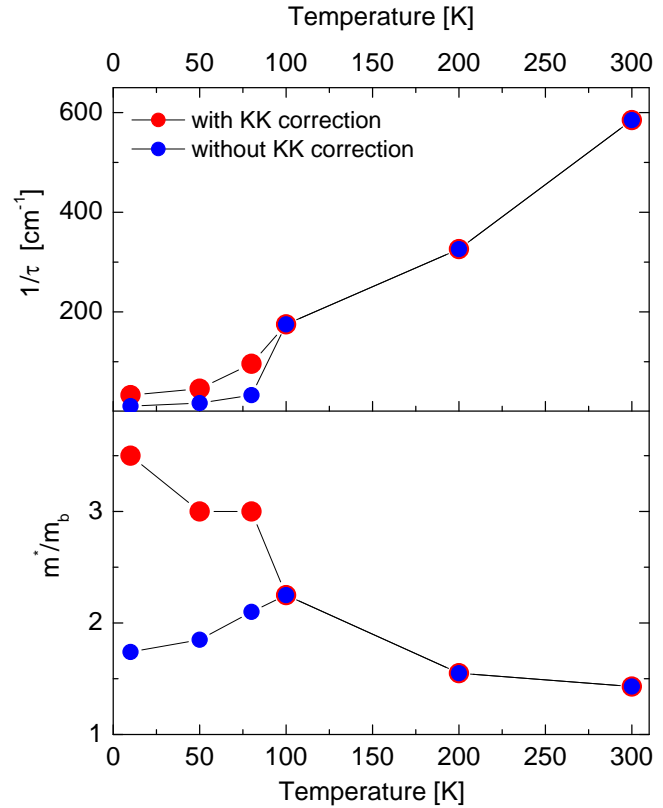}}%
\vspace*{0.0cm}%
\caption{(Color online). (a) Temperature dependence of scattering rate 1/$\tau(\omega)$
before and after Kramers-Kronig correction.
(b) Temperature dependence of effective mass m$^*$($\omega$)/m$_b$ before and after
Kramers-Kronig correction. These values are extracted as explained in the text.
The effective mass without Kramers-Kronig correction appears to decrease in the superconducting
state. However, with Kramers-Kronig correction the effective mass continues to increase below T$_c$,
which indicates that the quasiparticles are more correlated than in the normal state.}
\vspace*{0.0cm}%
\label{fig:temp}
\end{figure}

In Fig.~\ref{fig:temp} we display the temperature dependence of
quasiparticle scattering rate and effective mass, both before (blue
circles) and after (red circles) Kramers-Kronig correction. The
values of scattering rate were extracted as the average values at
around 30 cm$^{-1}$. On the other hand the values of the mass were
obtained from the linear fits of the low frequency M$_1$($\omega$)
spectra (Eq.~\ref{eq:memory}). This method has proven to be more
reliable then a simple extrapolation, in particular in the normal
state. The temperature dependence of scattering rate is similar
before and after the correction, however the suppression of
1/$\tau$($\omega$) below T$_c$ is much less pronounced after the
correction. On the other hand, the effective mass reveals
dramatically different behaviour. Before Kramers-Kronig correction
the mass decreases in the superconducting, as anticipated above.
However, once the superfluid and the true quasiparticle properties
are exposed, we can see that the effective mass actually increases
as temperature decreases below T$_c$. This indicates that the
correlations are getting stronger in the superconducting state.


\section{Summary}
\label{sec:summary}

We presented a way of calculating intrinsic quasiparticle
properties, such as memory function and effective mass, in the
superconducting state. The procedure allows access to the
quasiparticle properties that were previously inaccessible to
infrared spectroscopy. The method was first tested on model data,
and then applied to infrared data on optimally doped Bi2212. The
calculations have revealed that intrinsic quasiparticle scattering
rate and effective mass are enhanced in the superconducting state.
In particular, the effective mass increases below T$_c$ compared
with the normal state values. This indicates that thermally excited
quasiparticle in the superconducting state are more correlated than
in the normal state.

We expect the method described here to be a useful tool for
infrared spectroscopy, which will allow quasiparticle properties to
be studied for the first time in the superconducting state. In
particular, there are several issues in the cuprates that can be
immediately addressed using this new procedure. Scaling analysis
\cite{iman13,homes13} can now be extended below T$_c$, and the
question of Fermi vs non-Fermi liquid quasiparticles can be studied.
Another important issue is the controversy regarding the doping
dependence of the quasiparticle effective mass. Namely it was
recently show using quantum oscillations \cite{sebastian10} that
the quasiparticle effective mass in YBa$_2$Cu$_3$O$_{6+x}$ in the
superconducting state diverges as doping is reduced. This finding
is in apparent contradiction with previous infrared measurements
\cite{padilla05}. Using Hall data to discriminate between
carrier density {\it n} and effective mass m$^*$ contributions
to infrared spectral weight, Padilla {\it et al.} \cite{padilla05}
found that the effective mass in both YBa$_2$Cu$_3$O$_{6+x}$ and
La$_{2-x}$Sr$_x$CuO$_4$ was constant across the phase diagram.
On the other hand, by fitting a strong-coupling expressions
in the normal state, van Heumen {\it et al.} \cite{heumen09} arrived
at a factor of two decrease of the mass enhancement factor when the
doping is increased from 0.1 to 0.21 holes per CuO$_2$ unit, in
agreement with the behaviour predicted from dynamical mean field
theory \cite{haule07}. Using the procedure outlined in this
paper one can now access quasiparticle effective mass below T$_c$
and address this important issue in the zero temperature limit.

\section{acknowledgments}

This work was supported by the Swiss National Science Foundation (SNSF)
through grant 200020-140761.

%
%

\begin{references}
\bibitem{superfluid-comm} In the cuprates, the superfluid density is not as
    high as in conventional supercondoctors, resulting in $\sigma_2(\omega) >
    \sigma_1(\omega)$ instead of $\sigma_2(\omega) \gg \sigma_1(\omega)$. See
    for example Fig.~\ref{fig:sigma}.
\bibitem{comm-dirk} It was shown by Molegraaf et al. \cite{molegraaf02}
that a fraction of spectral weight is transfered from the visible to the
infrared spectral range. Note however that this effect is below the error bars
of our calculations.
\end{references}

\end{document}